%% file: VSP.tex
\documentclass[aps, prl, twocolumn, 10pt, nofootinbib]{revtex4-1}
\usepackage{amsmath,amssymb,amsfonts}
\usepackage{graphicx}          
\usepackage{physics}
\usepackage{caption}
\usepackage{subcaption}
\usepackage{siunitx}
\usepackage{tikz}
\usepackage{hyperref}
\usepackage{bm}
\usepackage{float}
\usepackage{adjustbox}
\usepackage{multirow}

\usetikzlibrary{calc,positioning}
\usetikzlibrary{arrows}
\usetikzlibrary{decorations.pathmorphing,calc,shapes,shapes.geometric,patterns}
\newlength\PPP
\renewcommand{\vec}{\bm}
\renewcommand{\d}{\text{d}}

\renewcommand{\(}{\left(}
\renewcommand{\)}{\right)}
\renewcommand{\[}{\left[}
\renewcommand{\]}{\right]}

\begin{document}
\title{Preparing Narrow Velocity Distributions for Quantum Memories in Room-Temperature Alkali Vapours}
\author{D. Main$^{1}$, T. M. Hird$^{1,2}$,  S. Gao$^{1}$, E. Oguz$^{1}$, D. J. Saunders$^{1}$, I. A. Walmsley$^{1,3}$, P. M. Ledingham$^{1,4}$}
\email{P.Ledingham@soton.ac.uk}
\affiliation{$^1$Clarendon Laboratory, University of Oxford, Parks Road, Oxford, OX1 3PU, UK\\
$^2$Department of Physics and Astronomy, University College London, London WC1E 6BT, UK\\
$^3$QOLS, Department of Physics, Imperial College London, London SW7 2BW, UK\\
$^4$Department of Physics and Astronomy, University of Southampton, Southampton SO17 1BJ, UK}
\date{\today}
	
\begin{abstract}
Quantum memories are a crucial technology for enabling large-scale quantum networks through synchronisation of probabilistic operations. Such networks impose strict requirements on quantum memory, such as storage time, retrieval efficiency, bandwidth, and scalability. On- and off-resonant ladder protocols on warm atomic vapour platforms are promising candidates, combining efficient high-bandwidth operation with low-noise on-demand retrieval. However, their storage time is severely limited by motion-induced dephasing caused by the broad velocity distribution of atoms comprising the vapour. In this paper, we demonstrate velocity selective optical pumping to overcome this decoherence mechanism. This will increase the achievable memory storage time of vapour memories. This technique can also be used for preparing arbitrarily shaped absorption profiles, for instance, preparing an atomic frequency comb absorption feature.
\end{abstract}
	
\maketitle

Large-scale quantum networks are a means to distribute entanglement across many spatially separated nodes. Such networks provide unique functionality arising from their quantum character, such as communication with guaranteed security \cite{Gisin2007}, secure remote access to cloud quantum computers \cite{Wehner2018} and precision metrology and accurate clock synchronisation for navigation \cite{Komar2014}. Photons provide the means to inter-connect disparate nodes; their high capacity for encoding information, robustness to decoherence, and ability to exhibit quantum features in ambient conditions make them the ideal carriers of quantum information in quantum networks. However, lossy channels and non-deterministic operations prohibit network scaling. Quantum memories, coherent light-matter interfaces that store and recall quantum states of light, can multiplex over operations to synchronise e.g. outputs of multiple heralded photon sources \cite{Nunn2013}, entanglement generation in a repeater \cite{Sangouard2008}, as well as provide an platform for single-photon nonlinearities \cite{Distante2016} and optimal filtering of quantum photonic states \cite{Gao2019}. These capabilities signify the importance of quantum memories as a building block for future quantum networks (see \cite{Heshami2016} for a comprehensive review of the quantum memory state of the art).

\begin{figure}
\begin{subfigure}{0.3\linewidth}
\includegraphics{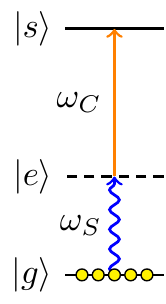}
\caption{}
\label{readin}
\end{subfigure}%
\begin{subfigure}{0.3\linewidth}
\includegraphics{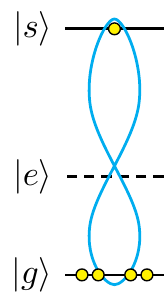}
\caption{}
\label{coherence}
\end{subfigure}%
\begin{subfigure}{0.3\linewidth}
\includegraphics{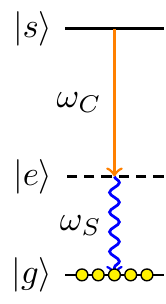}
\caption{}
\label{readout}
\end{subfigure}
\caption{Outline of a ladder memory protocol \cite{Kaczmarek2018,Finkelstein2018}: (a) A signal ($\omega_s$ blue) mode and a control ($\omega_c$ orange) mode are on two-photon resonance $\ket{g}-\ket{s}$ while being detuned $\Delta$ from the intermediate level $\ket{e}$ ($\Delta = 0$ shown here); (b) the signal field is mapped into an atomic coherence between $\ket{g}$ and $\ket{s}$; (c) the mapping is reversed via application of the control field allowing retrieval of the signal field.}\label{fig:ladder}
\end{figure}

For local quantum processing with demand on high clock rates, quantum memories with broadband acceptance are a requirement. On- and off-resonant $\Lambda$/ladder protocols on warm alkali vapour platforms are a promising candidate, in particular, the off-resonant cascaded absorption (ORCA) memory \cite{Kaczmarek2018} and the fast ladder memory (FLAME)\cite{Finkelstein2018}. These inherently noise-free protocols involve the mapping of a weak signal field into an atomic coherence between a ground state, $\ket{g}$, and a storage state, $\ket{s}$, via a two-photon absorption process in an atomic ensemble mediated by a strong control field, see Figure \ref{fig:ladder}. Kaczmarek et al. used the ORCA memory in caesium to store and recall heralded single photons, preserving quantum correlations upon read-out proving unequivocally that the protocol is noise free. 

\begin{figure*}
\begin{subfigure}{0.3\linewidth}
\includegraphics{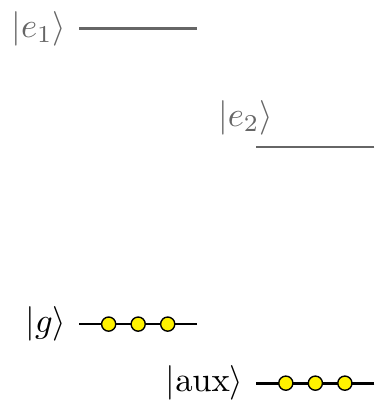}
\caption{}
\label{initial}
\end{subfigure}%
\begin{subfigure}{0.3\linewidth}
\includegraphics{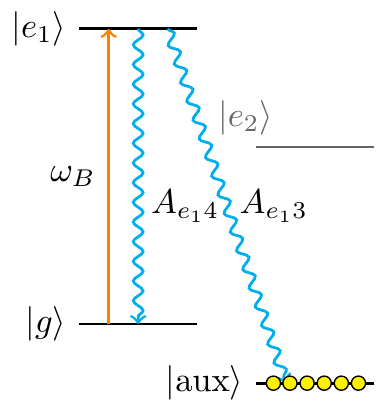}
\caption{}
\label{pump}
\end{subfigure}%
\begin{subfigure}{0.3\linewidth}
\includegraphics{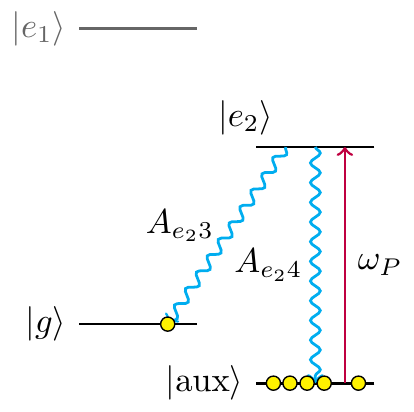}
\caption{}
\label{pump-back}
\end{subfigure}
\caption{Outline of the velocity selective pumping: (a) initial population in thermal equilibrium, (b) pump stage in which all population in the $\ket{g}$ state is transferred to the $\ket{\textrm{aux}}$ level, (c) velocity selective pumping re-pumps a velocity class back into the $\ket{g}$ hyperfine level for participation in the memory protocol. For the implementation in caesium vapour, we take the states $\ket{\textrm{aux}}$ and $\ket{g}$ to be the $F=3$ and $F=4$ hyperfine levels in the $6$~${}^2S_{1/2}$ manifold respectively.}\label{VelocitySelection}
\end{figure*}

The major challenge to overcome with this platform is  the motion-induced dephasing, which limited the storage time to $\SI{5.4}{ns}$ in \cite{Kaczmarek2018}, well below the storage state (in this case the $^6\mathrm{D}_{5/2}$) spontaneous emission lifetime limit ($\SI{60}{ns}$ \cite{Safronova2016}). Motion-induced dephasing is an inhomogeneous broadening mechanism that arises as a result of the Doppler effect and the thermal motion of atoms comprising the vapour. After the read-in process in ORCA at time $t = 0$, the signal field is mapped into a collective excitation over an atomic ensemble of $N$ atoms, described by $\ket{\psi\(t = 0\)} = \frac{1}{\sqrt{N}} \sum_{j=1}^Ne^{i\vec{k}_r\cdot\vec{r}_j\(0\)}\ket{g_1,\dots,s_j,\dots,g_N}$ where the magnitude of $\vec{k}_r$ is $k_r = \(\omega_S-\omega_C\)/c$ representing the wavevector of the atomic coherence created, with $\omega_{S/C}$ the angular frequency of the signal/control, and $\vec{r}_j$ the position vector of the $j^{\textrm{th}}$ atom in the ensemble. The read-out of the atomic coherence back to an optical field happens at a user-defined time $t$ with the $j^{\textrm{th}}$ atom moving to $\vec{r}_j\(t\) = \vec{r}_j\(0\)+\vec{v}_jt$ where $\vec{v}_j$ is the velocity vector of the $j^{\textrm{th}}$ atom. The atomic coherence at time $t$ is then $\ket{\psi\(t\)} = \frac{1}{\sqrt{N}}\sum_{j=1}^Ne^{i\vec{k}_r\cdot\vec{r}_j\(0\)}e^{i\vec{k}_r\cdot\vec{v}_jt}\ket{g,\dots,s_j,\dots,g}$. Assessing the overlap of these two atomic states:
\begin{equation}
\bra{\psi\(0\)}\ket{\psi\(t\)} = \frac{1}{N} \sum_{j=1}^Ne^{i\vec{k}_r\cdot\vec{v}_jt} \rightarrow \int \d v_zf\(v_z\)e^{i k_r v_zt}\label{coherence},
\end{equation}
where in the last step we consider propagation of the beams along the z-axis only so  $\vec{k}_r\cdot\vec{v} = {k_r}{v_z}$, and $f\(v_z\)$ is the velocity distribution of the atomic ensemble. This represents the Fourier transform of the velocity distribution, the characteristic spread of this distribution $\delta v$ relates to the characteristic spread in the conjugate space through the Fourier uncertainty principle, meaning the coherence dephases at a rate $\delta =  {k_r}{\delta v}$, or equivalently, the timescale on which the coherence degrades is $\tau_D = {1}/{{k_r}{\delta v}}$. When using the Raman quantum memory on a three-level system in a $\Lambda$ configuration in a warm vapour \cite{Nunn2013}, $k_r$ is small leading to dephasing times much longer (e.g. $\sim\SI{20}{\mu s}$ in caesium) than the atomic transit time. In the ladder configuration that we are considering here, $k_r$ is now much larger, leading to a $\SI{5.4}{ns}$ storage time for the $6S_{1/2}$-$6P_{3/2}$-$6D_{5/2}$ ladder in caesium \cite{Kaczmarek2018} and a $\SI{100}{ns}$ storage time for the $5S_{1/2}$-$5P_{3/2}$-$5D_{5/2}$ ladder in rubidium \cite{Finkelstein2018}.

Common approaches to negate this effect of Doppler dephasing are to work at degeneracy, i.e. $\omega_S=\omega_C$ such that $k_r = 0$, or to reduce the spread of velocities in the distribution such that $\delta v = 0$. The former approach would introduce unwanted complications to the memory protocol regarding the filtering of the signal from the control and the control inducing spurious atomic excitations into the storage state, while the latter approach typically involves atom trapping techniques resulting in added experimental complexity compared to vapour cells. Another method demonstrated by Finkelstein et al \cite{Finkelstein20} whereby inhomogeneous dephasing is eliminated by dressing the collective state to an auxillary state which doesn't suffer from this decoherence process. The added complication here is that an additional high-power laser is required to dress the collective state with some auxiliary sensor state, the wavelength and power of which depends on the level used.

Velocity selective optical pumping is a technique that allows preparation of atoms in a warm vapour such that only atoms of a specific velocity class are in the memory ground state $\ket{g}$ and the rest of the atoms are populating an auxiliary state $\ket{\textrm{aux}}$ that does not take part in the memory protocol. This results in a much narrower velocity distribution of atoms participating in the protocol, reducing $\delta v$ and thereby increasing $\tau_D$, while still being implemented in a warm vapour cell. This process is exactly spectral hole-burning, which is often used in rare-earth-ion doped media as a means to prepare the absorbing line for various quantum memory protocols, such as the atomic frequency comb \cite{DeRiedmatten2008}, controlled reversible inhomogeneous broadening \cite{Hedges2010}, electromagnetically induced transparency \cite{Heinze2013} and the spectral-hole memory protocol \cite{Kutluer2016}.

The approach of velocity selective pumping in warm vapour has been previously used to generate steep atomic dispersion features for slow light demonstrations \cite{Akulshin08, Camacho2006}, narrowband atomic-line frequency filters \cite{Cere2009}, and in high-resolution spectroscopy \cite{Marian2004, Aumiler2005, Ban2006, Vujicic2007}. Rubio et al. \cite{Rubio2018} propose creating a velocity selected atomic frequency comb via piece-wise adiabatic passage to transfer velocity classes between states in Barium vapour.

In this paper we present numerical simulations of the rate equations describing velocity-selective optical pumping for the specific case of caesium vapour at room temperature, which we verify with an experimental realisation.

\section{Theory}
In this section we outline the velocity-selective optical pumping technique and present a rate equation analysis model of the technique. Our approach to velocity-selective optical pumping is performed in two steps, depicted in Figure \ref{VelocitySelection}. The first step involves optically pumping the entire Doppler-broadened population out of the ground state, $\ket{g}$, and into a long-lived state $\ket{\textrm{aux}}$ that neither participates in the memory protocol nor rapidly decays back to the ground state. This initial step is referred to as the \textit{pump}. The second step, referred to as the \textit{pump-back}, makes use of a narrow-band laser ($\sim\textrm{MHz}$) to re-pump a narrow population from $\ket{\textrm{aux}}$ to $\ket{g}$, that is, a narrow velocity distribution into the state $\ket{g}$ depending on the frequency width and power of the \textit{pump-back} laser. We model this in caesium vapour where we take the states $\ket{g}$ and $\ket{\textrm{aux}}$ to be the the $F=4$ and $F = 3$ hyperfine levels in the $6$~${}^2S_{1/2}$ manifold.

We define $n_3\(v_z\)$ and $n_4\(v_z\)$ to be the number of atoms per unit volume in a particular velocity class populating the $F = 3$ and $F = 4$ hyperfine levels respectively. The pump step, shown in Figure \ref{pump}, involves transferring population in the $F = 4$ hyperfine level to an excited state $\ket{e_1}$ in the $6{}^2P_{3/2}$ manifold which then decays spontaneously to the hyperfine levels in the $6$~${}^2S_{1/2}$ manifold. The pump-back step, shown in Figure \ref{pump-back} is the reverse process, transferring a narrow spread of velocities from the $F = 3$ to the $F = 4$ levels via an excited level $\ket{e_2}$. The rate equations for the population inversion density, $n^*\(v_z\) = n_4\(v_z\) - n_3\(v_z\)$, and the excited population density $n_{e_1}\(v_z\)$ throughout the pump and pump-back steps are given by equations
\begin{align}
\dv{n^*}{t} =  &\mp \(n_F - \frac{g_F}{g_{e_j}}n_{e_j}\)\frac{B_{Fe_j}}{c}\nonumber\\
&\times \int_0^\infty I_k\(\omega\)g_L\(\omega-\omega_0\[1+\frac{v_z}{c}\]\)\d\omega\nonumber\\
&+ n_{e_j}\(A_{e_j4}-A_{e_j3}\)\label{rateEq1}\\
\dv{n_{e_j}}{t} =& \(n_F - \frac{g_F}{g_{e_j}}n_{e_j}\)\frac{B_{Fe_j}}{c}\nonumber\\
&\times \int_0^\infty I_k\(\omega\)g_L\(\omega-\omega_0\[1+\frac{v_z}{c}\]\)\d\omega\nonumber\\
&-n_{e_j}\(A_{e_j4}+A_{e_j3}\)\label{rateEq2}
\end{align}
where the subscripts  $F=\{4,3\}$, $j = \{1,2\}$ label the ground and excited states respectively, and $k=\{\textrm{pump},\textrm{pump-back}\}$ labels the pump mode. The integral in \ref{rateEq1} describes the interaction of the populations with the pump (pump-back) mode which is a narrowband pulse-carved CW laser centered on frequency $\omega_P$ ($\omega_{PB}$) with spectral intensity $I_P\(\omega\)$ ($I_{PB}\(\omega\)$). The lifetime-broadened lineshapes for each velocity class with Doppler-shifted resonant frequency $\omega_0\(1+v_z/c\)$ are given by $g_L$. $A_{e_f F}$ and $B_{Fe_j}$ are the Einstein A \& B coefficients describing spontaneous decay from $\ket{e_j}$ to $\ket{F}$ and the stimulated transfer from the hyperfine state $\ket{F}$ to the excited state $\ket{e_j}$ respectively. The second term therefore represents spontaneous decay of population from the excited states.

\begin{figure}
\includegraphics[width=0.7\linewidth]{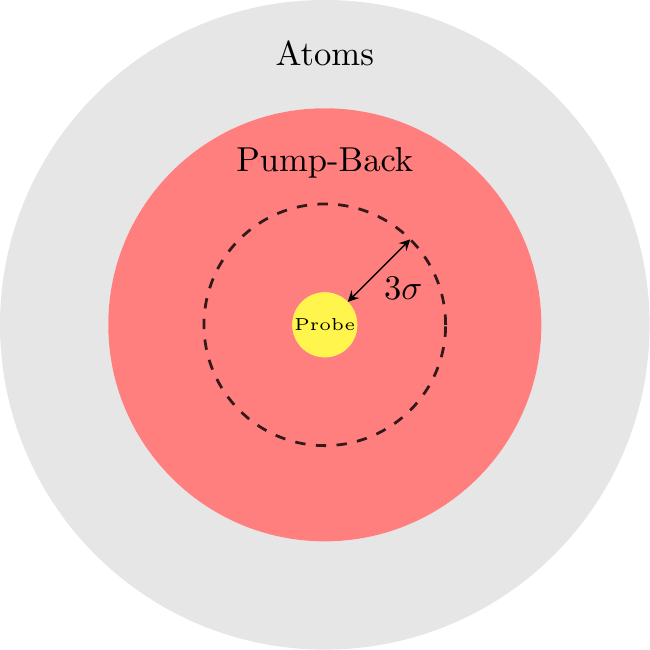}
\caption{Cross-section of the caesium sample showing the arrangement of the pump-back and probe modes. The beam waist of the pump-back (probe) mode is \SI{4}{\milli\meter} (\SI{0.3}{\milli\meter}). We show a $3\sigma$ length-scale that corresponds to the distance from the probe mode an atom would travel in $2\mu$s while moving with a radial velocity of three standard deviations of the velocity distribution ($3 v_{th}$). At $~\SI{23}{\celsius}$ this corresponds to a distance of \SI{0.82}{\milli\meter}.}\label{spatial_regions}
\end{figure}

In the above model, we have made the simplifying assumption that the atoms are confined to a spatially uniform area within the pump-back beam throughout the velocity selection process. Figure \ref{spatial_regions} shows a cross-section of the caesium sample, outlining the arrangement of the pump-back and probe modes. The $3\sigma$ length-scale (\SI{0.82}{\milli\meter}) shown in the figure corresponds to the distance that an atom moving at three standard deviations of the longitudinal velocity distribution could travel from the probe mode in \SI{2}{\micro\second}. We therefore assume that after velocity selection, the atoms within the probe mode have been sufficiently isolated from the significantly in-homogeneous area surrounding the approximately homogeneous $3\sigma$ region. Using a Maxwell-Boltzmann velocity distribution, we estimate that the timescale over which the movement of atoms between different spatial regions is $\sim\SI{7}{\micro\second}$, corresponding to a drift rate of $\gamma_D\sim1.4\times\SI{e5}{\per\second}$.

For the spectroscopy, we consider a weak scanning probe on the $D2$ transition. The optical cross sections for the transitions $6{}^2S_{1/2}$~$\(F=3,4\)$ to $6{}^2P_{3/2}$~$\(F'=2,3,4,5\)$ given a probe frequency $\omega$ are
\begin{equation}
\sigma_{4\rightarrow F'}\(\omega,v_z\) = B_{FF'}\frac{\hbar\omega}{c}g_L\(\omega-\omega_{FF'}\[1+\frac{v_z}{c}\]\)
\end{equation}
where $B_{FF'}$ are the Einstein $B$ coefficients, $g_L$ are the lifetime-broadened lineshapes of the transitions, and $\omega_{FF'}$ are the resonant frequencies of the transitions. In the weak probe limit the transmission, $T\(\omega\)$, of the scanning probe through the caesium cell is modelled as
\begin{equation}\label{transmission}
-\ln T\(\omega\) = L\sum_{\(F,F'\)}\int_{-\infty}^\infty n_F\(v_z\)\sigma_{F\rightarrow F'}\(\omega,v_z\)\d v_z
\end{equation}
where $L$ is the length of the caesium cell and the sum is over pairs of hyperfine levels $\(F,F'\)$ that obey the electric dipole selection rules $\Delta F=0,\pm1$. This describes a sum of the convolutions of the velocity distributions, $n_F\(v_z\)$, with the non-Doppler shifted cross-sections $\sigma_{F\rightarrow F'}\(\omega,0\)$.

\section{Experiment}

Velocity selective optical pumping was implemented using the experimental setup shown in Figure \ref{experiment}. A narrowband, continuous-wave (CW) laser tuned to the $F = 4\rightarrow F'=4$ transition in the $D_2$ line ($\SI{852}{nm}$ - Sacher ECDL) was used to perform the \textit{pump} and a second CW-laser tuned to the $F = 3\rightarrow F'=4$ transition in the $D_1$ line ($\SI{895}{nm}$ - Toptica DFB) was used to perform the \textit{pump-back}. A narrowband, CW scanning \textit{probe} was used to perform absorption spectroscopy on the $F = 4\rightarrow F' = 3,4,5$ transitions in the $D_2$ line (Toptica DL Pro). Acousto-optic modulators (AOMs) were used to carve pulses out of the CW-lasers. The pump and the pump-back were combined after the AOMs using a dichroic mirror before a telescope increased the size of the beams to $\SI{1.5}{mm}$. A polarising beam splitter was used to transmit the pumping modes and reflect the counter-propagating probe which was then detected by a silicon avalanche photodetector (Thorlabs APD120). The pickoff in the probe mode allows to calibrate changes in power associated with the scanning frequency. The caesium cell has length \SI{2.5}{\centi\meter} and radius \SI{1}{\centi\meter}. The cell has a paraffin coating which plays a key roll in preventing spin flipping upon cell wall collisions while redistributing the atomic velocities such that the initial pumping becomes efficient even with a narrowband laser \cite{Li2017a,Li2017b}. 

\begin{figure}
\begin{subfigure}{\linewidth}
\includegraphics[width=\linewidth]{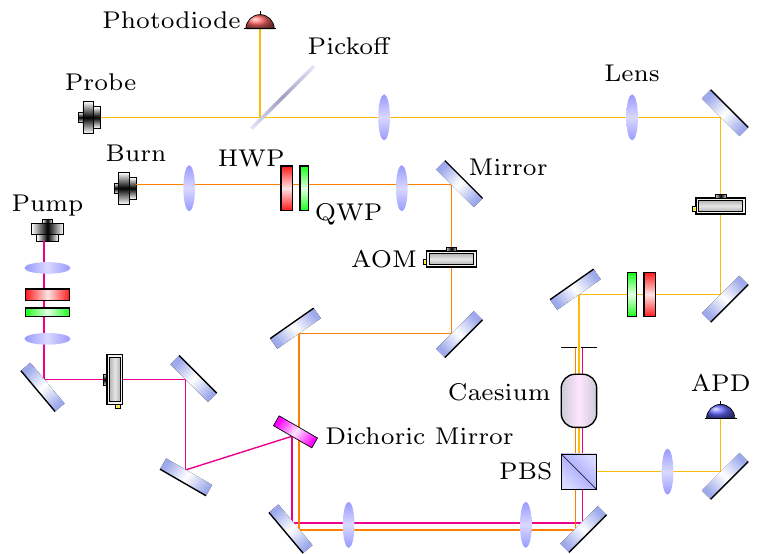}
\caption{}
\label{experiment}
\end{subfigure}\newline\vspace{0.3cm}%
\begin{subfigure}{\linewidth}
\includegraphics[width=\linewidth]{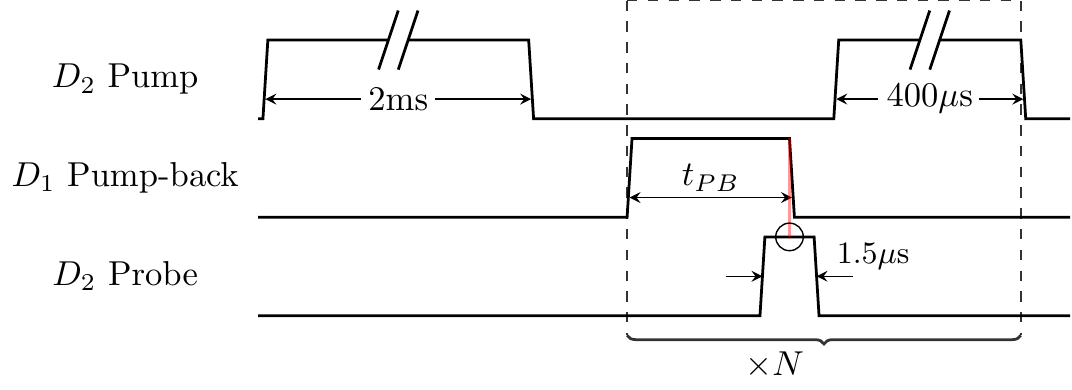}
\caption{}
\label{experiment}
\end{subfigure}
\caption{(a) Experimental setup for investigating the velocity selective optical pumping. Three CW-lasers -- probe, pump, and pump-back -- were pulsed using AOMs. A dichoric mirror combined the pump and the pump-back whilst a PBS directed the counter-propagating probe onto an APD to perform spectroscopy. 
\\(b) Pulse sequence used in the preparation and spectroscopy of the velocity selected atomic population. An initial pump of length \SI{2}{ms} on the $D_2$ transition prepares the atoms in the ground $F = 3$ hyperfine state. After this preparation, a repeating sequence of pump-back, probe, and reset pump is employed $N$ times. The probe is positioned such that its midpoint is aligned with the point at which the pump-back is turned off. This allows us to probe the transmission spectrum as soon as velocity selection has been performed.}
\end{figure}

\begin{figure*}
\centering
\begin{tikzpicture}

\node[inner sep=0pt] (img) at (0,0)
    {\resizebox{15cm}{!}{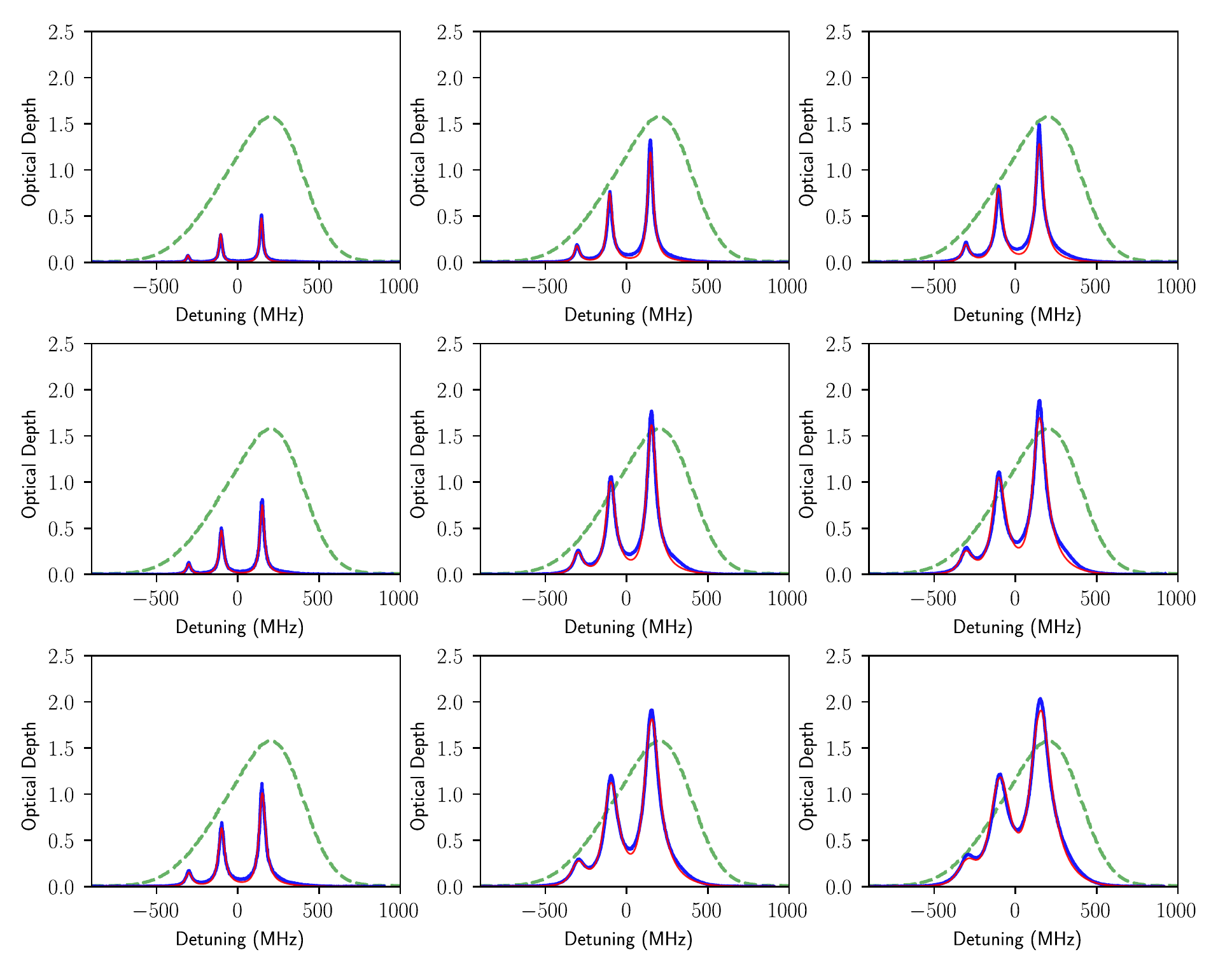}};

\node[draw, single arrow, minimum height=10.92cm, minimum width=7mm, single arrow head extend=2mm, anchor=west] at (-5.5cm, 6.4cm) {Increasing Pump-Back Time};
\node[draw, single arrow, minimum height=8.58cm, minimum width=7mm, single arrow head extend=2mm, anchor=west,rotate = -90] at (-8,4.5) {Increasing Pump-Back Power};
\end{tikzpicture}
\caption{Velocity-selective pumping spectra. The blue lines are the optical depths of the $D_2$ $F=4\rightarrow F'=3,4,5$ transitions as a function of frequency for various pump-back times and powers at $\SI{23}{\celsius}$. The horizontal axis is the detuning from the transition between the $F=4$ hyperfine level in the $6{}^2S_{1/2}$ manifold and the $6{}^2P_{3/2}$ manifold. From top to bottom, the pump-back powers are: \SI{0.86}{mW,} \SI{4.1}{mW}, and \SI{10.5}{mW}. From left to right, the pump-back times are: \SI{0.2}{\mu s}, \SI{1.2}{\mu s}, and \SI{2}{\mu s}. The red lines are the simulated spectra. The green line is the unpumped, Doppler broadened spectrum.}\label{Spectra}
\end{figure*}

Since the drift rate of the atoms ($\gamma_D \sim 10^5\SI{}{Hz}$) is much greater than the probe scanning rate (\SI{5}{Hz}) it is not possible to probe the full spectrum after a given velocity-selective pumping experiment. We instead piece a full spectrum together with around a thousand probe measurements. The sequence begins with the pump mode open for approximately \SI{2}{ms} and power of $\SI{20}{mW}$, emptying the $F = 4$ hyperfine level. This is followed by 20 modules of (1) pump-back mode on for a duration of $t_P$ and varying power level; (2)  frequency scanning probe mode on for $\SI{1.5}{\mu s}$ at a power level orders of magnitude below saturation, where we take the level on the photodiode immediately as the pump mode has ended; (3) a reset pump for duration $\SI{400}{\mu s}$ which re-empties the $F = 4$ hyperfine level. The AOMs have a switching time of $\SI{100}{ns}$. This sequence prepares and probes a velocity selected feature around 20 times per probe scan, each probe measurement being at a different frequency, and we repeat this sequence 50 times piecing together 1000 probe measurements to complete the full spectrum.

To assess the depth and width of the measured features, we fit simulated spectra  calculated from equation \ref{transmission} using population densities obtained by solving numerically the rate equations \ref{rateEq1} and \ref{rateEq2} that describe the velocity selection. The pump-back power and linewidth, and selected velocity class were used as fitting parameters. The nominal values for the linewidth and selected velocity class were $\sim$\SI{6}{\mega\hertz} and $\sim$\SI{-100}{\meter\per\second}, respectively. These parameters were allowed to vary from their nominal values to account for relaxation dynamics not captured by our model. For lower pump-back powers and times, the fitted parameters were in good agreement with the nominal values. However at higher pump-back powers and times, the fitted power varied by up to $50\%$ of the nominal value and the fitted linewidth increased up to as high \SI{70}{\mega\hertz}. By using equation \ref{coherence} and the velocity distribution $f\(v_z\)\propto n_4\(v_z\)$, we obtain an estimate for the dephasing timescale as the $1/e$ time for the mod-square of the overlap of the atomic states. Note that this timescale relates only to the motion-induced dephasing, the spontaneous lifetime of the storage state used in the memory protocol will also inhibit the achievable memory lifetime.  

Finally, as a way to characterise the performance of the anti-relaxation coating, velocity selection was performed then left to relax. The time-dependence of the peak absorption during relaxation was obtained and indicated the presence of two exponential decaying terms of opposite signs \cite{Li2017,Graf2005},
\begin{equation}
T\(t\) = A_1e^{-\gamma_s t}-A_2e^{-\gamma_f t}+c
\end{equation}
where $A_1,A_2>0$ and $c$ are constants, $\gamma_s$ is the slow decay rate, and $\gamma_f$ is the fast decay rate. The fast decay rate was determined to be $\gamma_f=8(2)\times10^4\text{s}^{-1}$, whereas the slow decay rate was determined to be $\gamma_s=40(7)\text{s}^{-1}$. By comparison to the decay rate predicted by the Maxwell-Boltzmann distribution, we attribute the fast decay to the movement of atoms between the different spatial regions in the cell (see Figure \ref{spatial_regions}), and attribute the slow decay to the thermalisation of the atomic vapour.

\section{Discussion}
The measured spectra and the fitted simulated spectra, a sample of which are presented in Figure \ref{Spectra}, show very good qualitative agreement between our model and measurement. We identify power broadening as the main mechanism affecting the spectral shape of the features. For the case of low pump-back power and duration, the features are Lorentzian. As the pump-back power is increased, power broadening occurs as the central velocity class is saturated and more atoms from wings of the pump-back spectrum address a wider range of velocity class. At large pump-back powers and times, the features are almost washed out as a result of interaction of the pump-back with the majority of the velocity classes. Another mechanism that we would expect to be significant at longer pump-back times is the drifting of the atoms; velocity selected atoms leave the probe region, collide with the paraffin coated wall and re-enter within a different velocity class. However, over the range of pump-back times explored in this experiment, we expect this mechanism to be negligible. Due to the $T^{1/2}$ scaling of the thermal velocity, we expect this effect to be accentuated at higher temperatures.

\begin{figure}[h!]

\centering
\includegraphics[width=\linewidth]{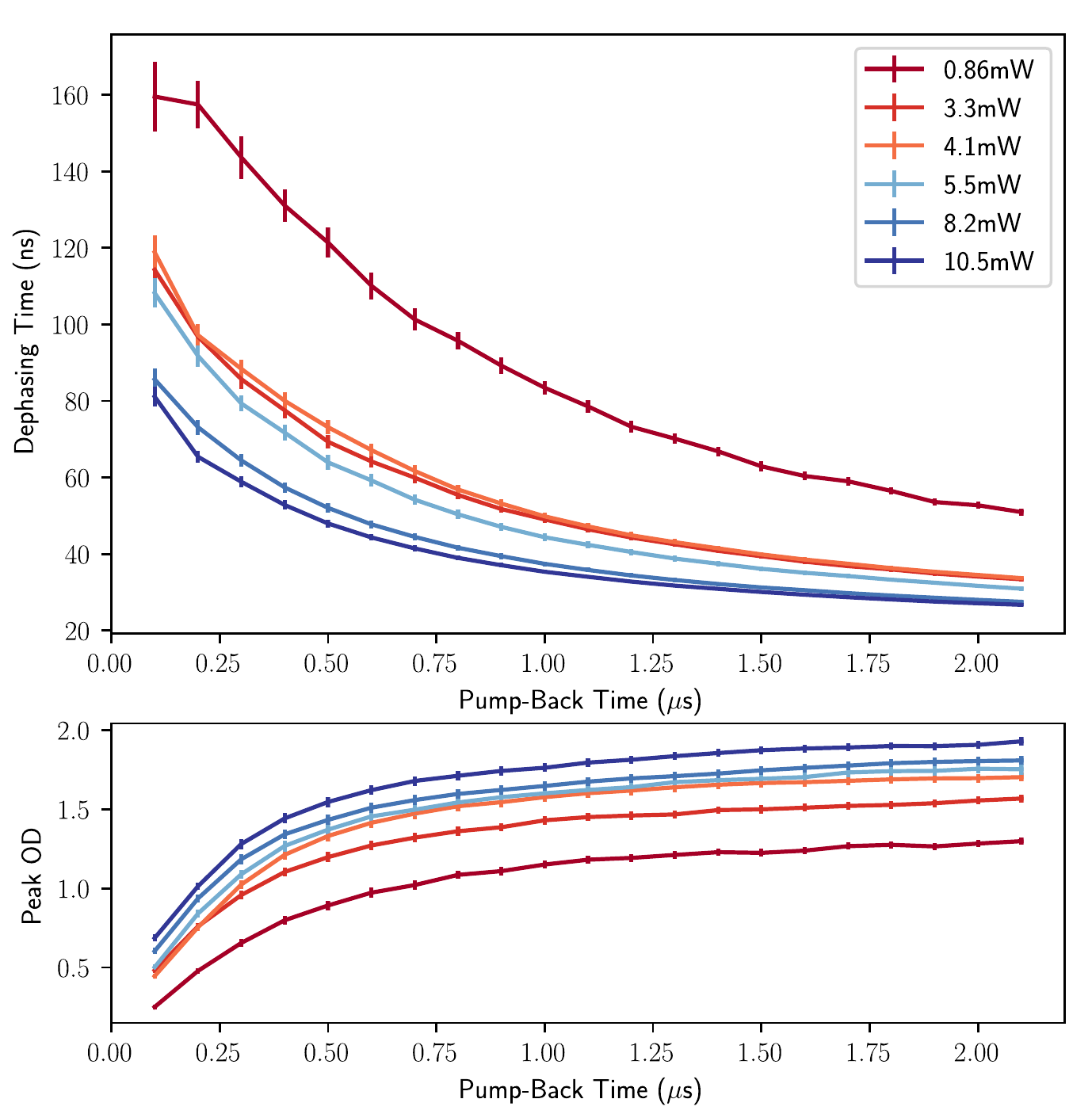}
\caption{Dephasing times (top) and peak optical depths (OD) (bottom) at different pump-back powers as a function of pump-back time for $\SI{23}{\celsius}$.\label{deph}}
\end{figure}

Adapting our simulation as a fitting tool to our measurements allowed extraction of a velocity distribution, $n_4\(v_z\)$, which was used for the direct calculation of the motion-induced dephasing timescale. Figure \ref{deph} shows the pump-time dependence of this timescale at different powers. 
The estimated lower bound of the dephasing time is \SI{14}{\nano\second} in the absence of velocity selection. From Figure \ref{deph} it is clear that increasing the optical depth will incur a trade-off in dephasing time; application of velocity selection in a memory will have to take this compromise into account.

Although the velocity selection was not performed at the temperature at which the ORCA memory has been implemented ($\SI{90}{\celsius}$), the results from this project indicate the potential of the technique. The motion-induced dephasing times in Figure \ref{deph} range from $30$ns to $160$ns, exceeding the spontaneous lifetime of the storage state used in ORCA is $60$ns. It is therefore possible to use velocity selection to bring the memory into a regime in which spontaneous emission is the dominant decoherence mechanism, rather than motion-induced dephasing.

We now consider the benefit that velocity selection could offer to the ORCA memory protocol. Firstly as applied to the demonstrated operation in \cite{Kaczmarek2018} before considering a broader scope of applications. Since the Doppler decoherence rate $\delta = k_r\delta v$ we may also consider ladder transitions that would otherwise give shorter lifetimes than to be useful due to the wave-vector mismatch. We therefore consider how velocity selective pumping could enhance the memory lifetimes of ORCA protocols implemented in caesium and rubidium, making use of the simulation presented earlier to predict the coherence as a function of storage time with and without velocity selection, the results of which are presented in Table \ref{tab:predicted_coherence}. The enhancement factor, $\beta$, defined to be the ratio between the predicted memory lifetimes with and without application of velocity selective pumping, is included in order to highlight the potential enhancement offered by the technique. These transitions were chosen to illustrate the advantage of the scheme could bring to the longest lived memory lifetimes (917nm/776nm) and how it may be applied to telecom transitions within the atomic species (1470nm/1528nm).

\begin{table}[h]
\renewcommand{\arraystretch}{1.5}
\begin{tabular}{c||c|c|c|c|c}
& Upper Level & Wavelength & No VSP & VSP & $\beta$\\\hline
\multirow{2}{*}{Cs}  & $6D_{5/2}$  & \SI{917}{\nano\meter}  & \SI{12.5}{\nano\second} & \SI{47.4}{\nano\second}  & 3.8\\ \cline{2-6} 
& $7S_{1/2}$  & \SI{1470}{\nano\meter} & \SI{2.3}{\nano\second}  & \SI{19.0}{\nano\second} &  8.3 \\ \hline
\multirow{2}{*}{Rb} & $5D_{5/2}$  & \SI{776}{\nano\meter} & \SI{97.9}{\nano\second} & \SI{217.4}{\nano\second} & 2.2\\ \cline{2-6} 
& $4D_{5/2}$  & \SI{1528}{\nano\meter} & \SI{1.4}{\nano\second}  & \SI{18.9}{\nano\second} & 13.5
\end{tabular}
\caption{
Predicted $1/e$ timescales for the memory efficiency of ORCA memory protocols implemented in Caesium and Rubidium, with and without the application of velocity selective pumping (VSP). In each case we consider two potential upper levels that could be used to form the ladder memory. The enhancement factor, $\beta$, is defined as the ratio between the timescales with and without velocity selective pumping.
}\label{tab:predicted_coherence}
\end{table}

\vspace{1cm}

For each of the considered cases, the pump-back power and time were taken to be \SI{1}{mW} and \SI{0.1}{\micro\second}. We modelled selecting the zero-velocity class and took the linewidth of the pump-back laser to be \SI{6}{\mega\hertz}. For the purposes of these comparisons we consider the lifetime to be dependent on only the Doppler dephasing and the lifetime of the excited state; other lesser, effects such as atoms drifting out of the beam and hyperfine beating  \cite{Kaczmarek2018} are neglected. Table \ref{tab:predicted_coherence} predicts a significant enhancement for ORCA memory lifetime (characterised by $\beta$) due to application of velocity selective pumping. In particular, memories utilising a telecom transition stand to receive a potential order of magnitude enhancement through application of this technique as a result of resolving the Doppler dephasing caused by the large wavevector mismatches involved. The ability to interface telecom photons with an ORCA memory that has an enhanced memory lifetime offers the possibility for large scale quantum networks.

\section{Conclusion}

In conclusion we have proposed and demonstrated a method by which we are able to selectively pump a single velocity class for as a means for reducing the inhomogenous dephasing. We apply this this method in caesium vapour and compare our experimental results to numeric simulation. Finally we comment on how this relates to the ORCA memory protocol and the potential lifetime enhancement it offers, particularly in the future of warm vapour quantum memories with telecom photons. 

The authors would like to thank X.  Peng and  H.  Guo  for  kindly  providing  the  paraffin-coated cell for this work. This work was supported by the UK Engineering and Physical Sciences Research Council (EPSRC) through the Standard Grant No. EP/J000051/1, Programme Grant No. EP/K034480/1, the EPSRC Hub for Networked Quantum Information Technologies (NQIT), and an ERC Advanced Grant (MOQUACINO). TMH is supported via the EPSRC Training and Skills Hub InQuBATE Grant EP/P510270/1.

\bibliographystyle{unsrt}

\end{document}

%% file: spectra_grid2.pdf_tex
\begingroup%
  \makeatletter%
  \providecommand\color[2][]{%
    \errmessage{(Inkscape) Color is used for the text in Inkscape, but the package 'color.sty' is not loaded}%
    \renewcommand\color[2][]{}%
  }%
  \providecommand\transparent[1]{%
    \errmessage{(Inkscape) Transparency is used (non-zero) for the text in Inkscape, but the package 'transparent.sty' is not loaded}%
    \renewcommand\transparent[1]{}%
  }%
  \providecommand\rotatebox[2]{#2}%
  \ifx\svgwidth\undefined%
    \setlength{\unitlength}{557.39612849bp}%
    \ifx\svgscale\undefined%
      \relax%
    \else%
      \setlength{\unitlength}{\unitlength * \real{\svgscale}}%
    \fi%
  \else%
    \setlength{\unitlength}{\svgwidth}%
  \fi%
  \global\let\svgwidth\undefined%
  \global\let\svgscale\undefined%
  \makeatother%
  \begin{picture}(1,0.78243739)%
    \put(0,0){\includegraphics[width=\unitlength,page=1]{spectra_grid2.pdf}}%
  \end{picture}%
\endgroup%